\documentclass[12pt, prd, showpacs]{revtex4}
\usepackage{amssymb}
\usepackage{amsmath}

\input{tcilatex}

\begin{document}

\title{Ultra-high energy collisions in static space-times: single versus
multi-black hole cases}
\author{O. B. Zaslavskii}
\affiliation{Department of Physics and Technology, Kharkov V.N. Karazin National
University, 4 Svoboda Square, Kharkov 61022, Ukraine}
\affiliation{Institute of Mathematics and Mechanics, Kazan Federal University, 18
Kremlyovskaya St., Kazan 420008, Russia}
\email{zaslav@ukr.net }

\begin{abstract}
We consider collision of two particles near static electrically charged
extremal black holes and elucidate the conditions under which the energy in
the centre of mass frame $E_{c.m.}$ grows unbounded. For a single black
hole, we generalize the results obtained earlier for the Reissner-Nordstr%
\"{o}m metric, to distorted black holes. In the multi-black hole space-time,
qualitatively new features appear. If the point of collision is close to at
least two horizons simultaneously, unbounded $E_{c.m.}$ are possible (i)
without fine-tuning of particles' parameters, (ii) for an arbitrary mutual
orientation of two velocities. Such a combination of properties (i) and (ii)
has no analogues in the single black hole case and facilitates the condition
of getting unbounded $E_{c.m.}$. Collisions in the electro-vacuum
Majumdar-Papapetrou metric (several extremal black holes in equilibrium) is
analyzed explicitly.
\end{abstract}

\maketitle

\section{Introduction}

High energy processes near black holes are of important both from the
theoretical and astrophysical points of view. Now, the interest to such
processes increased significantly after findings of Ba\~{n}ados, Silk and
West (hereafter, BSW). They found that the energy $E_{c.m.}$ in the centre
of mass frame of two particles colliding near the extremal Kerr black hole,
can grow unbounded \cite{ban}. Meanwhile, high energy collision near
rotating black holes were considered before in a series of papers \cite{pir1}
- \cite{pir3}. There is a crucial difference, however, between these papers
and \cite{ban}. In the first case, unbounded $E_{c.m.}$ $\ $is achieved when
both colliding particles move in the same direction towards a black hole.
This is possible if one of colliding particles has fine-tuned parameters. In
the situation discussed in \cite{pir1} - \cite{pir3}, the effect is due to
head-on collision and does not require any fine-tuning. (There is also an
intermediate case when collision occurs on the near-horizon circular orbit.)
However, another complication arises here. As one particle in the immediate
vicinity of the horizon should move not towards a black hole but away from
it, this requires special conditions - say, necessity of preceding
collisions (multiple scattering scenario). In both situations \cite{ban} and 
\cite{pir1} - \cite{pir3}, it is implied that a single black hole is present.

Quite recently, new set-up was suggested in \cite{di}. In that paper,
collisions in the multi-black hole metric were studied. More specifically,
the exact Majumdar-Papapetrou solution of electrovacuum Einstein equations 
\cite{mj}, \cite{pet} was exploited with the concrete analysis of geodesic
particle motion in the background of two black holes. Unbounded $E_{c.m.}$
was obtained for extremely small space separation between black holes. In
doing so, different cases were compared in \cite{di} to the BSW effect near
a single black hole depending on the sign of radial velocities (whether
particles move in the same direction or in the opposite one).

The goal of the present work is to show that unbounded $E_{c.m.}$ can be
achieved near multi-black holes for collision of \textit{any} two particles
moving in the vicinity of multi-black hole metric (starting from the two
black hole configuration). The angle $\psi $ between their velocity can be
arbitrary, in the case $\psi =0$ or $\psi =\pi $ we return to the situation
considered in \cite{di}. This degree of freedom has crucial consequences
since, as will be clear below, it enables to arrange high-energy collision
without fine-tuning typical of the BSW effect \cite{prd}. And, multiple
scattering is not required now to have a particle moving away from a black
hole. The fact that important constraints on getting such energies are
relaxed, enlarges chances that the effects under discussion can have (at
least, in principle) observational relevance.

We suggest a unified picture for the cases of multi-black and single black
holes and give full classification of cases when unbounded $E_{c.m.}$ are
possible for charged distorted extremal black holes.

The paper is organized as follows. In Sec. II we consider a generic single
electrically charged extremal black hole without assumption of spherical
symmetry. Equations of motion are listed, the conservation law is formulated
in terms of the spatial velocity. In Sec. III we analyze particle collisions
in this background and classify the possible cases depending on whether or
not unbounded $E_{c.m.}$ are possible. In Sec. IV we discuss the multi-black
hole metric using the Majumdar - Papapetrou solution as an example. In Sec.
V, general situation for the case of two black holes is discussed. Two kinds
of limiting transitions in revealed in Sec. VI, brief comparison with
previous results on this subject \cite{di} is made. In Sec. VII, we discuss
the problem of collisions from a general viewpoint analyzing kinematic
underlying factors that lead to unbounded $E_{c.m.}$ Summary of main results
is given in Sec. VIII.

Throughout the paper we use units in which fundamental constants are $G=c=1$.

\section{Single black hole: basic equations}

Let us consider the generic metric of a static black hole. It can be written
in the Gauss normal coordinate system (that always exists at least in some
vicinity of the horizon):

\begin{equation}
ds^{2}=-N^{2}dt^{2}+dn^{2}+g_{AB}dx^{A}dx^{B},  \label{met}
\end{equation}%
where $A$, $B=2,3$ and all coefficients do not depend on $t$. The horizon
lies at $N=0$. Let us consider equations of motion for test particles. The
energy $E=-P_{t}$ is conserved due to staticity. Here, $P_{\mu }$ is the
generalized momentum. Let us suppose that the system is electrically
charged. Then,%
\begin{equation}
mu_{t}=P_{t}+q\varphi \text{,}
\end{equation}%
$\varphi $ is the electric potential, $q$ being the particle's charge, $m$
the particle's mass, $u^{\mu }$ the four-velocity. The equations of motion
read%
\begin{equation}
m\dot{t}=\frac{X}{N^{2}}\text{,}  \label{0}
\end{equation}%
\begin{equation}
X=E-q\varphi \text{,}  \label{x}
\end{equation}%
\begin{equation}
m\dot{n}=\varepsilon \frac{Z}{N},  \label{1}
\end{equation}%
\begin{equation}
Z=\sqrt{X^{2}-N^{2}(m^{2}+g_{AB}u^{A}u^{B})}\text{.}  \label{z}
\end{equation}%
Dot denotes derivative with respect to the proper time $\tau $. The
parameter $\varepsilon =-1$ if a particle moves towards a black hole and $%
\varepsilon =+1$ in the opposite case. To derive (\ref{1}), we used the
normalization condition 
\begin{equation}
u_{\mu }u^{\mu }=-1.  \label{norm}
\end{equation}

To get insight into kinematics of motion, it is instructive to introduce the
orthogonal tetrad basis $h_{(a)\mu }$ ($a$ runs from $0$ to 3$)$ and define 
\begin{equation}
V_{(i)}=V^{(i)}=-\frac{u^{\mu }h_{\mu (i)}}{u^{\mu }h_{\mu (0)}}  \label{v1}
\end{equation}%
(its counterpart for rotating black holes is analyzed in Sec. III of \cite%
{72}), $i=1,2,3$.

For the metric (\ref{met}), it is natural to introduce the tetrad according
to%
\begin{equation}
h_{(0)\mu }=-N(1,0,0,0)  \label{h0}
\end{equation}%
\begin{equation}
h_{(1)\mu }=(0,1,0,0),
\end{equation}%
\begin{equation}
h_{(A)\mu }=(0,0,s_{(A)2},s_{(A)3})\text{,}  \label{h3}
\end{equation}

where $x^{\mu }=(t,n,x^{2},x^{3})$, $a=2,3$, $s_{(A)c}s_{(B)}^{c}=\delta
_{AB}$, where $c=2,3$ and $\delta _{AB}$ is the Kronecker symbol. It follows
from (\ref{0}), (\ref{1}), (\ref{v1}) that%
\begin{equation}
V^{(1)}=\frac{\dot{n}}{\dot{t}N}=\varepsilon \frac{Z}{X}\text{,}  \label{vn}
\end{equation}%
\begin{equation}
V^{(A)}=\frac{\dot{x}^{b}s_{(A)b}N}{X}.  \label{va}
\end{equation}

One obtains from (\ref{vn}), (\ref{va}) that 
\begin{equation}
X=\frac{mN}{\sqrt{1-V^{2}}}=mN\gamma _{1}\text{.}  \label{xn}
\end{equation}%
Here,%
\begin{equation}
V^{2}=V^{(1)2}+V^{(2)2}+V^{(3)2}\text{, }
\end{equation}%
the individual gamma-factor 
\begin{equation}
\gamma _{1}=\frac{1}{\sqrt{1-V_{1}^{2}}},
\end{equation}%
\begin{equation}
\gamma _{1}=\frac{X_{1}}{m_{1}N}.
\end{equation}

Eq. (\ref{xn}) coincides with eq. (29) of \cite{k} but is valid in a more
general situation, without assumption about spherical symmetry.

\section{Collisions near single black hole}

Let us consider collision between two particles whose characteristics are
labeled by indices 1 and 2. One can define the energy in the centre of mass
frame $E_{c.m.}$ according to%
\begin{equation}
E_{c.m.}^{2}=-(m_{1}u_{1}^{\mu }+m_{2}u_{2}^{\mu })(m_{1}u_{1\mu
}+m_{2}u_{2\mu })=m_{1}^{2}+m_{2}^{2}+2m_{1}m_{2}\gamma \text{,}
\end{equation}%
where 
\begin{equation}
\gamma =-u_{1\mu }u_{2}^{\mu }  \label{ga}
\end{equation}%
has the meaning of the Lorentz factor of relative motion. Direct
calculations gives us from (\ref{0}), (\ref{1})%
\begin{equation}
\gamma =\frac{X_{1}X_{2}-\varepsilon _{1}\varepsilon _{2}Z_{1}Z_{2}}{%
m_{1}m_{2}N^{2}}-g_{AB}u_{1}^{A}u_{2}^{B}\text{.}  \label{ga1}
\end{equation}

In general, we cannot solve equations of motion explicitly. Fortunately,
this is not required in the given context. We only assume that $%
g_{AB}u_{1}^{A}u_{2}^{B}$ remains finite on the horizon. In the spherically
symmetric space-time and for radial motion, $g_{AB}u_{1}^{A}u_{2}^{B}=0$.
For the Reissner-Nordstr\"{o}m metric, we return to the case considered in 
\cite{jl}.

Now, we want to analyze (\ref{ga1}) thus generalizing the results \cite{jl}
to the distorted black holes when spherical symmetry is not required. In
what follows, we call a particle critical, if $X_{H}=0$, and usual if $%
X_{H}\neq 0$. If $X_{H}\neq 0$ but is extremely small, we call a particle
near-critical. Here, subscript "H" denotes the quantity calculated on the
horizon. Actually, $X_{H}>0$ for usual particles due to the forward in time
condition $\dot{t}>0$.

Let us consider the vicinity of the horizon. Then, for small $N$, we have
for a usual particle the expansion%
\begin{equation}
Z=X_{H}-\frac{N^{2}(m^{2}+g_{AB}u^{A}u^{B})_{H}}{2X_{H}}+O(N^{4}).  \label{u}
\end{equation}%
It is also seen from (\ref{vn}), (\ref{va}) that in the horizon limit the
velocity is directed along the normal to the horizon, so $V^{(3)}=O(N)\ll
V^{(1)}\approx \varepsilon $. Thus for the angle $\phi _{0}$ between the
velocity and the normal to the horizon we have 
\begin{equation}
\cos \phi _{0}=\pm 1  \label{1-1}
\end{equation}%
depending on the direction of motion.

From now on, we assume that black holes under considerations are extremal.
This enables to avoid complications connected with the fact that the
critical particle cannot reach the horizon of a nonextremal black hole \cite%
{prd}. For the critical particle, assuming the validity of the Taylor
expansion, we can write%
\begin{equation}
X=CN+DN^{2}+...  \label{xc}
\end{equation}

Then,%
\begin{equation}
Z\approx N\sqrt{C^{2}-m^{2}-(g_{AB}u_{1}^{A}u_{2}^{B})_{H}}\text{.}
\label{c}
\end{equation}

It follows from (\ref{vn}), (\ref{va}) that both components of the velocity $%
V^{(1)}$ and $V^{(3)}$ are, generally speaking, separated from zero and have
the same order, so the angle between the velocity and the normal to the
horizon $\phi _{0}$ can be arbitrary depending on particle's characteristics.

Now, we will consider collision between particles in two cases separately.

\subsection{Analogue of BSW effect, $\protect\varepsilon _{1}\protect%
\varepsilon _{2}=+1$}

If collision occurs between two usual or two critical particles, it follows
from (\ref{ga1}) that $\gamma $ remains finite on the horizon. The only case
of interest is when, say, particle 1 is critical and particle 2 is usual.
Then,%
\begin{equation}
\gamma \approx \frac{\left( X_{2}\right) _{H}(C_{1}-\sqrt{%
C_{1}^{2}-(m_{1}^{2}+g_{AB}u_{1}^{A}u_{1}^{B})_{H}})}{N}  \label{uc1}
\end{equation}%
is unbounded that represents just the analogue of the BSW effect.

\subsection{Head-on collisions, $\protect\varepsilon _{1}\protect\varepsilon %
_{2}=-1$}

Now, two particles move in the opposite radial directions. For collisions of
two usual particles near the horizon,%
\begin{equation}
\gamma \approx \frac{2\left( X_{1}\right) _{H}\left( X_{2}\right) _{H}}{%
m_{1}m_{2}N^{2}}\text{.}  \label{uu}
\end{equation}%
Thus if $N$ is small enough, $\gamma $ can become as large as one likes.

If critical particle 1 collides with a usual one 2,%
\begin{equation}
\gamma \approx \frac{2C_{1}\left( X_{2}\right) _{H}}{m_{1}m_{2}N}\text{.}
\label{uc}
\end{equation}

The Lorentz factor grows more slowly but it diverges on the horizon anyway.
For collision of two critical particles, $\gamma $ remains finite.

Thus we can enumerate all possible configurations and the results for $%
\gamma $ in Table 1. Here, "u -c" means collision between a usual and the
critical particles, etc. 

\begin{tabular}{|l|l|l|l|}
\hline
& opposite directions, $\varepsilon _{1}\varepsilon _{2}=-1$ & arbitrary
different directions & coinciding directions, $\varepsilon _{1}\varepsilon
_{2}=+1$ \\ \hline
$u-u$ & $N^{-2}$ & impossible & finite \\ \hline
$u-c$ & $N^{-1}$ & $N^{-1}$ & $N^{-1}$ \\ \hline
$c-c$ & finite & finite & finite \\ \hline
\end{tabular}

Table 1. Possible cases of particle collisions near the horizon of a single
black hole.

\section{Majumdar-Papapetrou systems}

Now we will consider high energy collision in the background of the Majumdar
- Papapetrou solution \cite{mj}, \cite{pet}. Let we have $n$ extremal black
holes in equilibrium.\ Then,%
\begin{equation}
ds^{2}=-U^{-2}dt^{2}+U^{2}(dx^{2}+dy^{2}+dz^{2})\text{,}  \label{mp}
\end{equation}

\begin{equation}
U=N^{-1}=1+\dsum\limits_{i=1}^{n}\frac{M_{i}}{l_{i}}\text{, }l_{i}=\sqrt{%
(x-x_{i})^{2}+(y-y_{i})^{2}+(z-z_{i})^{2}}\text{,}  \label{un}
\end{equation}%
$M_{i}$ is a mass of i-th black hole. Each black hole has the electric
charge $Q_{i}=M_{i}$. The "points" ($x_{i},y_{i}$, $z_{i})$ are actually not
the points in a usual sense but correspond to the horizons of a finite area 
\cite{h}.

The electrostatic potential%
\begin{equation}
\varphi =1-U^{-1}=1-N\text{,}  \label{pot}
\end{equation}%
where the constant is chosen to ensure $\varphi \rightarrow 0$ at infinity.
It follows from (\ref{x}) and (\ref{pot}) that%
\begin{equation}
X_{1,2}=E_{1,2}-q_{1,2}+q_{1,2}N\text{.}  \label{xmp}
\end{equation}

It is worth noting that, as the potential $\varphi =1$ on each horizon, the
criticality condition $X_{H}=0$ reduces to $E=q$. Such a relation involves
the characteristics of a particle itself only and is the same for all
horizons. This simplifies greatly the analysis of particle collisions (see
below). If, say, particle 1 is critical, we have 
\begin{equation}
X_{1}=\frac{q_{1}}{U}  \label{1cr}
\end{equation}%
that agrees with (\ref{xc}), where now only the first term is nonzero.

Now, the natural choice of a tetrad is slightly different from (\ref{h0}) - (%
\ref{h3}):%
\begin{equation}
h_{(0)}^{\mu }=U(1,0,0,0)\text{, }h_{(0)\mu }=-U^{-1}(1,0,0,0)\text{,}
\label{0h}
\end{equation}%
\begin{equation}
h_{(1)}^{\mu }=U^{-1}(0,1,0,0)\text{, }h_{(1)\mu }=U(0,1,0,0)  \label{1h}
\end{equation}%
and similarly for $h_{(2)}^{\mu }$, $h_{(3)}^{\mu }$, where $x^{\mu
}=(t,x,y,z)$. Then, the tetrad components of a velocity%
\begin{equation}
V_{(i)}=V^{(i)}=\frac{m\dot{x}^{i}}{X}\text{,}  \label{v}
\end{equation}%
\begin{equation}
V^{2}=1-\frac{m^{2}}{X^{2}U^{2}}.  \label{V}
\end{equation}%
If two particles collide, simple calculation of (\ref{ga}) gives us 
\begin{equation}
\gamma =\gamma _{1}\gamma _{2}(1-\vec{V}_{1}\vec{V}_{2})\text{, }\gamma
_{1,2}=\frac{X_{1,2}}{m_{1,2}}U\text{,}  \label{ga12}
\end{equation}%
where the scalar product 
\begin{equation}
\vec{V}_{1}\vec{V}_{2}=\sqrt{1-\frac{m_{1}^{2}}{X_{1}^{2}U^{2}}}\sqrt{1-%
\frac{m_{2}^{2}}{X_{2}^{2}U^{2}}}\cos \psi  \label{v12}
\end{equation}%
is calculated in the flat Euclidean space, $\psi $ being the angle between $%
\vec{V}_{1}$ and $\vec{V}_{2}$ in the point of collision.

The key moment consists in the appearance of the factor $\cos \psi $, where $%
\psi $ is the angle between three-velocities of particles in the point of
collision. In the case of a single black hole, any usual particle approaches
the horizon along the normal to it according to (\ref{1-1}), so $\cos \psi
=\pm 1$. By contrast, now this a free parameter. This is quite natural since
if there are several black holes, a particle cannot have a velocity that is
perpendicular to all of them, even if separation between different black
holes is small.

Let collision between two particles occur just in the region where
separation between different black holes is small, so $U$ is large, $%
N=U^{-1} $ is small. The analysis of $\gamma $ goes similarly to the case of
a single black hole and we obtain the following table of possible situations.

\begin{tabular}{|l|l|l|}
\hline
& $\gamma $ for different directions, $\psi \neq 0$ & $\gamma $ for
coinciding directions, $\psi =0$ \\ \hline
$u-u$ & $N^{-2}$ & finite \\ \hline
$u-c$ & $N^{-1}$ & $N^{-1}$ \\ \hline
$c-c$ & finite & finite \\ \hline
\end{tabular}

Table 2. Possible cases of particle collisions near multi-black hole for
small separation.

It is seen from (\ref{v12}) that the relative sign of velocities is
determined by $\cos \psi $. Let, say, two usual particles collide. Then, for 
$\psi =0$ (particles move in the same direction) and $\psi =\pi $ (particles
move in the opposite direction) the results presented in Table 2 reduce to
those in Table 1 for $\varepsilon _{1}\varepsilon _{2}=+1$ and $\varepsilon
_{1}\varepsilon =-1$, respectively.

Meanwhile, Table 2 contains some new variants to achieve unbound $\gamma $
that were impossible according to Table 1. Now, this becomes possible if two
usual particles or one usual and one critical particles move in different
(in particular, opposite) directions.

If the point of collision is much more close to one of black holes than to
others (say, to black hole 1), we can leave in the sum (\ref{un}) the
corresponding term only. Then, the problem is reduced to that considered in
the previous Section devoted to a single black hole.

\section{Two black holes: basic equations}

To illustrate the foregoing general features, let us consider the case of
two black holes of equal masses $M$ situated on the z-axis in the points $+a$
and $-a$. This is just the example discussed in \cite{di}. We compare the
approach of \cite{di} with ours and reveal that, actually, there are two
different kinds of limits that lead to high energy collisions.

It is instructive to make now transformation $x=\rho \cos \phi $, $y=\rho
\sin \phi $ to the cylindric coordinate system in which%
\begin{equation}
ds^{2}=-U^{-2}dt^{2}+U^{2}(d\rho ^{2}+\rho ^{2}d\phi ^{2}+dz^{2})\text{. }
\label{u2}
\end{equation}%
Here,%
\begin{equation}
U=1+\frac{M}{\sqrt{\rho ^{2}+(z-a)^{2}}}+\frac{M}{\sqrt{\rho ^{2}+(z+a)^{2}}}%
.
\end{equation}

In what follows, we restrict ourselves by motion in the equatorial plane $%
z=0 $, where%
\begin{equation}
U=1+\frac{2M}{\sqrt{\rho ^{2}+a^{2}}}\text{.}  \label{ura}
\end{equation}

\subsection{Motion with zero angular momenta}

To begin with, we consider the case when both particles move along straight
lines, so their angular momenta are equal to zero like in Sec. 4 of \cite{di}%
. However, we make emphasis on non-collinear motion in this case that
expands the set of possibilities how to obtain high $\gamma $ and $E_{c.m.}$.

It is easy to show that, if a particle starts its motion along x-axis, it
keeps moving along this direction, so $z=0=y$ \cite{di}. However, if we
rotate the axis at the angle $\psi $, nothing changes. Therefore, particle 2
can move along the direction $\phi =\psi =const$. We have from (\ref{norm})
and (\ref{0}) that%
\begin{equation}
\left( \frac{dx}{d\tau }\right) ^{2}+\left( \frac{dy}{d\tau }\right) ^{2}=%
\dot{\rho}^{2}+\rho ^{2}\dot{\phi}^{2}=\frac{X^{2}-m^{2}N^{2}}{m^{2}}.
\label{3}
\end{equation}%
Let two particle pass through the center $x=y=0=\rho $ and collide just in
this point. Then, $\phi =0$ for particle 1 and $\phi =\psi =const$ for
particle 2. Calculating the gamma factor (\ref{ga}) and taking into account (%
\ref{ga12}), (\ref{v12}), (\ref{3}) one obtains that%
\begin{equation}
\gamma =\frac{X_{1}X_{2}-Z_{1}Z_{2}\cos \psi }{m_{1}m_{2}}U^{2}\text{.}
\label{gp}
\end{equation}

In the point of collision,%
\begin{equation}
U=1+\frac{2M}{a}\text{.}
\end{equation}%
If $a\ll M$, we see that $U\gg 1$ and $N\ll 1$, so $\gamma $ can become
unbounded. The diversity of possibilities is just described by Table 2.
Collisions studied in \cite{di} correspond to $\psi =0$, $\psi =\pi $.

Thus we do not need to analyze the full equations of motion that are quite
cumbersome \cite{di}.\ Moreover, actually there is no necessity to
constraint motion by additional special conditions (like motion along the
line). This is because there are general formulas (\ref{ga12}), (\ref{v12})
that relate $\gamma ,$ characteristics of motion and the metric function $U$
from which the effect of unbounded $\gamma $ can be obtained if $U$ is big
enough.

\subsection{More general set-up}

Now, let a particle have the angular momentum $L$ and move in the equatorial
plane. Then, in the metric (\ref{u2}), its motion, by close analogy with (%
\ref{0}) - (\ref{z}), is described by equations%
\begin{equation}
m\dot{t}=XU^{2}\text{,}
\end{equation}%
\begin{equation}
m\dot{\phi}=\frac{L}{U^{2}\rho ^{2}}\text{,}
\end{equation}%
\begin{equation}
m\dot{\rho}=\varepsilon ZU\text{,}
\end{equation}%
\begin{equation}
Z=\sqrt{X^{2}-\frac{m^{2}}{U^{2}}-\frac{L^{2}}{U^{4}\rho ^{2}}}\text{.}
\label{zu}
\end{equation}

It is convenient to use the tetrads (now $x^{i}=t,\rho ,z,\phi )$%
\begin{equation}
h_{(0)\mu }=-U^{-1}(-1,0,0,0)\text{,}
\end{equation}%
\begin{equation}
h_{(1)\mu }=U(0,1,0,0)\text{,}
\end{equation}%
\begin{equation}
h_{(2)\mu }=U(0,0,1,0)\text{,}
\end{equation}%
\begin{equation}
h_{(2)\mu }=U(0,0,0,\rho )\text{.}
\end{equation}

The analogues of formulas (\ref{vn}), (\ref{va}) read%
\begin{equation}
V^{(1)}=V\cos \beta =\varepsilon \frac{Z}{X}\text{,}  \label{1v}
\end{equation}%
\begin{equation}
V^{(3)}=V\sin \beta =\frac{L}{\rho U^{2}X}\text{,}  \label{3v}
\end{equation}%
where $V$ is given by eq. (\ref{V}), $\beta $ is the angle characterizing
direction of motion of an individual particle.

For the relative angle between particles in the point of collision we have%
\begin{equation}
\cos \psi =(\varepsilon _{1}\varepsilon _{2}Z_{1}Z_{2}+\frac{L_{1}L_{2}}{%
U^{4}\rho ^{2}})\frac{1}{\sqrt{X_{1}^{2}-m_{1}^{2}U^{-2}}}\frac{1}{\sqrt{%
X_{2}^{2}-m_{2}^{2}U^{-2}}}\text{.}  \label{cos}
\end{equation}%
Direct calculation of the Lorentz factor of relative motion of two particles
(\ref{ga}) gives us%
\begin{equation}
\gamma =\gamma _{1}\gamma _{2}(1-\vec{V}_{1}\vec{V}_{2})=\frac{U^{2}}{%
m_{1}m_{2}}[(X_{1}X_{2}-\varepsilon _{1}\varepsilon _{2}Z_{1}Z_{2})-\frac{%
L_{1}L_{2}}{\rho ^{2}U^{4}}]\text{.}  \label{g2}
\end{equation}

Using (\ref{1v}), (\ref{3v}), (\ref{V}), one can also write (\ref{g2}) in
the form

\begin{equation}
\gamma =\frac{X_{1}X_{2}U^{2}}{m_{1}m\,_{2}}(1-\sqrt{1-\frac{m_{1}^{2}}{%
X_{1}^{2}U^{2}}}\sqrt{1-\frac{m_{2}^{2}}{X_{2}^{2}U^{2}}}\cos \psi ).
\label{gcos}
\end{equation}%
If, say, $L_{1}=0$ from the very beginning, so $\sin \beta _{1}=0$, one can
check using (\ref{zu}), (\ref{1v}) - (\ref{cos}) that eq. (\ref{g2}) is
reduced to eq. (\ref{gp}).

\section{Two kinds of limiting transitions}

We are interested in the behavior of $\gamma $ for small separation $a$ and
small $\rho =\rho _{0}$ in the point of collision, where one can expect
indefinitely large growth of $\gamma $. Correspondingly, there are two
relevant limits depending on what quantity is sent to zero first.

\subsection{Limit A, $\protect\gamma _{A}\equiv \lim_{a\rightarrow 0}\lim_{%
\protect\rho _{0}\rightarrow 0}\protect\gamma $}

Let two particles collide in the centre or very closely to it. This means
that in the point under consideration $\rho =\rho _{0}\rightarrow 0$, 
\begin{equation}
U^{2}\rho _{0}\rightarrow 0\text{.}  \label{A}
\end{equation}%
Then, the right hand side of (\ref{3v}) grows unbound both for the usual and
critical particles, if $L\neq 0$. Meanwhile, the left hand side remains
bounded, its value is less than 1. Therefore, for $L\neq 0$ the scenario
under discussion is impossible. However, we can arrange this scenario,
provided $L$ becomes smaller and smaller in this process, $L\sim \rho
_{0}\rightarrow 0$ \ This applies to each of two particles and to the angle
between them $\psi =\beta _{1}-\beta _{2}$ that remains arbitrary. If
separation $a$ between two black holes is small, $U(\rho =0)=1+\frac{2M}{a}$
is large. Then, we return to the case of collision for noncollinear motion
considered above - see Table 2 and Subsection V A that describes now the
behavior of $\gamma _{A}$.

\subsection{Limit B, $\protect\gamma _{B}\equiv \lim_{\protect\rho %
_{0}\rightarrow 0}\lim_{a\rightarrow 0}\protect\gamma $ }

Now, instead of (\ref{A}), the opposite case is realized:%
\begin{equation}
U^{2}\rho _{0}\rightarrow \infty .  \label{B}
\end{equation}

Now, $L_{1}$ and $L_{2}$ may be arbitrary. (But they cannot be zero
simultaneously. This would correspond to collision in the centre, so we
would return to case A instead of B.)

It follows from (\ref{3v}) that, for a usual particle, $\sin \beta
\rightarrow 0$. Taking into account (\ref{ura}) and (\ref{1cr}), one can see
that for the critical particle $\beta $ can be arbitrary. Therefore, the
angle $\psi $ between two usual particles is equal to $0$ or $\pi $ for two
particles and can be arbitrary if at least one particle is critical.

This is also seen from (\ref{cos}). Let $Z_{1.2}\neq 0$, Then, for collision
of two particles it follows from (\ref{zu}) that $Z\approx \sqrt{X^{2}-\frac{%
m^{2}}{U^{2}}},$ the second term in (\ref{cos}) \ in parentheses is
negligible as compared to the first one due to (\ref{B}). We obtain 
\begin{equation}
\cos \psi \approx \varepsilon _{1}\varepsilon _{2}\text{.}
\end{equation}

The Lorentz factor $\gamma $ is finite if $\varepsilon _{1}\varepsilon
_{2}=+1$ (motion in the same direction)$.$ If $\varepsilon _{1}\varepsilon
_{2}=-1$ (motion in the opposite direction), this factor behaves like%
\begin{equation}
\gamma \approx \frac{2X_{1}X_{2}}{m_{1}m_{2}}U^{2}
\end{equation}%
and becomes unbound.

In a similar way, one can analyze the rest of possible configurations with
the results that coincide with those presented in Table 1, provided $%
Z_{1,2}\neq 0$ in the point of collision.

\subsection{Special case: collision near turning point}

In the above treatment of eq, (\ref{cos}), the second term in  parentheses
was much smaller than the first one. The opposite situation arises if $%
Z_{1}(\rho _{0})=0$. Physically, it means that for particle 1, the collision
point coincides with the turning point. Then, it is seen from (\ref{1v})
that $\cos \beta _{1}=0$. It follows from (\ref{zu}), (\ref{cos}), (\ref{g2}%
) that now%
\begin{equation}
\cos \psi =\pm \frac{L_{2}}{U^{2}\rho \sqrt{X_{2}^{2}-m_{2}^{2}U^{-2}}}\text{%
,}
\end{equation}%
\begin{equation}
\gamma =\frac{1}{m_{1}}\frac{1}{m_{2}}[X_{1}X_{2}U^{2}-\frac{L_{1}L_{2}}{%
\rho ^{2}U^{2}}]\text{.}  \label{gc}
\end{equation}

It is seen from (\ref{ura}) that $\rho U$ is finite in the limit under
discussion. Correspondingly, it follows from (\ref{ura}) and (\ref{1cr})
that particle 1 is critical, $X_{1}\sim U^{-1}$. If particle 2 is also
critical, $X_{2}\sim U^{-1}$, so $\gamma $ is finite, $\psi $ can be
arbitrary. If particle 2 is usual, $\cos \psi \rightarrow 0$ due to (\ref{B}%
). Then, $X_{2}$ remains finite nonzero, $\gamma \sim U$ becomes unbound.

We see that the although collision in the turning point looks somewhat
different from other cases, the results completely fall in the general
scheme and are described by Table 1.

Formally, one can consider also limit C, for which%
\begin{equation}
\rho =\alpha a  \label{aa}
\end{equation}%
with $\alpha =O(1),$ so $\rho _{0}$ and $a$ tend to zero with the equal
rate. However, it is easy to check that limit C does not give any new as
compared to limit B, so Table 1 applies here.

\subsection{Comparison to previous studies}

In Ref. \cite{di} two situations were analyzed. In Sec. 4, collision with
zero angular momenta is considered in the centre $\rho =0$ that corresponds
to our limit A. In this sense, our results give generalization to the case
of an arbitrary angle $\psi $ between particles. In Sec. 7, collision
between particles with nonzero momenta was studied. Both particles were
taken to be identical, having angular momenta $L_{1,2}=\nu _{1,2}L$, where $%
\nu _{1,2}=\pm 1$, $L>0$. Different configurations of collision, depending
on $\varepsilon _{1}\varepsilon _{2}$ ($\beta _{1}\beta _{2}$ in notations
of \cite{di}) and $\nu _{1}\nu _{2}$ were analyzed in \cite{di} in detail
for different regions of parameters. Meanwhile, all of them are based on eq.
(38) that coincides with our eq. (\ref{aa}). Therefore, they correspond to
limit C that is equivalent to B, as is said above.

If $L=0$, both $L_{1}=L_{2}=0$. But this is impossible in case B or C, as is
explained above. Although some equations of Sec. 7 of \cite{di} do not
contain $L$ explicitly, $L$ is contained there implicitly. Say, in eq. (67)
of \cite{di}, $L$ is actually present through the parameter $\sigma $
defined in (63).

It is worth also noting that particles were assumed to be uncharged in \cite%
{di}, so both of them are usual. Our results represented in Table 2, include
usual and critical particles.

\subsection{Summary for collisions near two black holes}

To summarize the results of the present Section, there are two different
limits A and B (or, equivalently, C). For the corresponding regimes, two
opposite relations (\ref{A}) and (\ref{B}) hold. In the first case, the
angle $\psi $ between particles is arbitrary, $L_{1.2}\rightarrow 0$. In the
second one, momenta $L_{1,2}$ are arbitrary, the angle $\psi =0$ or $\pi $
for a usual particle or can be arbitrary for the critical one. It was stated
in Sec. 8 that there is a crucial difference from the BSW effect near a
single black hole. This is correct, but if one takes into account all
possible processes near a single black hole (including motion not only
towards a black hole), the whole set of possibilities for collisions near a
single black hole and two black holes (in variant B or C) is the same.
Therefore, case B (or C) does not give qualitatively new results from the
high energy collisions near a single black hole, even in spite of the
crucial difference in the geometries for such configurations. 

Meanwhile, case A has no analogues for collisions near a single black hole
at all.

Division to two different types of scenarios is connected with the high
symmetry of a system due to which angular momenta of particle are preserved,
there is a preferable direction of motion along the radius, etc. In a
general case, when black holes are situated irregularly, one cannot expect
analogues of a situation with a single hole in the region where gravitation
fields of different holes overlap, so that scenarios of type B is not
expected to be valid in general. Meanwhile, generalization of scenarios of
type A seem to retain their validity since they do not require any symmetry.
This issue is discussed in the next Section.

\section{General kinematic picture}

It is instructive to look at the problem from a more general viewpoint, not
restricting ourselves by the metric (\ref{mp}). Say, we can include matter
into consideration (dirty multi-black holes). Earlier, we showed that the
growth of $\gamma $ in the BSW effect, can be interpreted in terms of
relative motion \cite{k}. Below, we will relate directly $\gamma $ to $%
\gamma _{1}$ and $\gamma _{2}$ characterizing motion of each particle, to
extend consideration to a more general case of collision.

\subsection{Flat space-time}

First of all, let us consider collision of two particles 1 and 2 in the
simplest case of the flat space-time. The laboratory frame is labeled by
index 0. Taking, say, particle 1, we can use decomposition%
\begin{equation}
u_{1}^{\mu }=\gamma _{1}U^{\mu }+\beta _{1}n^{\mu },  \label{u1}
\end{equation}%
where $U^{\mu }$ is the four velocity of an observer attached to the
laboratory frame, $n^{\mu .}$is the vector orthogonal to it. One can obtain
from (\ref{u1}) directly that%
\begin{equation}
\gamma _{1}=-U_{\mu }u_{1}^{\mu }\text{.}
\end{equation}

It is seen that the quantity $\gamma _{1}$ has the meaning of the Lorentz
factor of motion of particle 1 with respect to "particle" 0. In other words,
this is just the gamma factor of particle 1 in the laboratory frame
(individual gamma-factor). It follows from the normalization conditions
\thinspace $U_{\mu }U^{\mu }=u_{\mu }u^{\mu }=-1$ that%
\begin{equation}
\beta _{1}^{2}=\gamma _{1}^{2}-1\text{.}
\end{equation}%
Calculating (\ref{ga}), we obtain%
\begin{equation}
\gamma =\gamma _{1}\gamma _{2}-\beta _{1}\beta _{2}\alpha \text{. }
\label{gb}
\end{equation}%
Here, $\alpha =n_{1\mu }n^{2\mu },\left\vert \alpha \right\vert <1.$

In the laboratory frame,%
\begin{equation}
u_{a}^{\mu }=\gamma _{a}(1,V_{a}\vec{n}_{a})\text{, }\alpha =\vec{n}_{1}\vec{%
n}_{2}\equiv \cos \psi \text{.}
\end{equation}%
\begin{equation}
\alpha V_{1}V_{2}=\vec{V}_{1}\vec{V}_{2}
\end{equation}%
\begin{equation}
\gamma _{1,2}=\frac{1}{\sqrt{1-V_{1,2}^{2}}}\text{.}  \label{g1}
\end{equation}

Eq. (\ref{gb}) can be rewritten as%
\begin{equation}
\gamma =\frac{1}{\sqrt{1-V^{2}}}=\gamma _{1}\gamma _{2}(1-V_{1}V_{2}\cos
\psi )=\gamma _{1}\gamma _{2}-a\sqrt{\gamma _{1}^{2}-1}\sqrt{\gamma
_{2}^{2}-1}\text{,}  \label{g12}
\end{equation}%
where $V$ is the relative velocity

In a slightly different form,%
\begin{equation}
\gamma =\gamma _{1}\gamma _{2}(1-\vec{V}_{1}\vec{V}_{2})\text{.}  \label{gf}
\end{equation}

Eqs. (\ref{g12}), (\ref{gf}) can be found, for example, in problem 1.3 of
the problem book \cite{teuk}.

Now, one can enumerate all possible cases.

1) If both $\gamma _{1}$ and $\gamma _{2}$ are finite, $\gamma $ is also
finite irrespective of the sign of $\alpha $.

2) Let $\gamma _{1}\gg 1$, $\gamma _{2}$ is finite. Then,%
\begin{equation}
\gamma \approx \gamma _{1}(1-\alpha \sqrt{\gamma _{2}^{2}-1})=\gamma
_{1}(1-\alpha V_{2})\gg 1
\end{equation}%
irrespective of the sign of $\alpha $. It means that the relative velocity
of particles, one of which moves with a speed separated from the speed of
light and the other one almost with the speed of light, is always close to
the speed of light.

3) Let $\gamma _{1}\gg 1$, $\gamma _{2}\gg 1$.

a) $\alpha \neq +1$. Then,%
\begin{equation}
\gamma \approx \gamma _{1}\gamma _{2}(1-\alpha )\gg 1
\end{equation}%
is unbounded.

b) $\alpha =+1$. 
\begin{equation}
\gamma \approx \frac{1}{2}(\frac{\gamma _{1}}{\gamma _{2}}+\frac{\gamma _{2}%
}{\gamma _{1}})\text{.}
\end{equation}

If $\gamma _{1}\sim \gamma _{2}$, the gamma factor $\gamma $ is finite. If $%
\gamma _{1}\gg \gamma _{2}$ or vice versa, $\gamma \gg 1$.

\subsection{Curved space-time}

One can introduce the tetrad orthogonal basis. Then, the components of
velocities should be understood according to (\ref{v1}). Previous formulas (%
\ref{g1}) - (\ref{gf}) retain their validity. For critical particle 1, $%
\gamma _{1}$ is finite. For usual particle 2, $\gamma _{2}\sim N^{-1}$.
There are two essential ingredients. (i) Collision occurs in the region
where $N\ll 1$, (ii) this is achieved due to small separation between black
holes. (ii) mutual orientation is arbitrary.\ 

Let both particles be usual, so $V_{1}\approx 1$ and $V_{2}\approx 1$, Then,
we have from (\ref{g12}) that%
\begin{equation}
\gamma \approx \gamma _{1}\gamma _{2}(1-\cos \psi )
\end{equation}%
is unbounded, provided $\psi \neq 0$. In doing so, $\gamma _{1}$ $\sim
\gamma _{2}\sim N^{-1}$, $\gamma \sim N^{-2}$.

If particle 1 is near-critical and particle 2 is usual,%
\begin{equation}
\gamma \approx \gamma _{1}\gamma _{2}(1-V_{1}\cos \psi )
\end{equation}

is also unbound. But now $\gamma _{1}$ is finite, $\gamma _{2}\sim \gamma
\sim N^{-1}$.

If both particles are near-critical, $\gamma _{1}$ and $\gamma _{2}$ are
finite, there is no effect at all.

\section{Conclusion}

Thus we gave full classification of possible scenarios of collisions in the
multi-black hole case. It unifies the BSW effect and head-on collisions in a
more general coherent picture. We saw that high-energy collisions near black
holes turn out to be "almost" universal phenomenon. There is a crucial
difference between high energy collisions near a single black hole and those
in the multi-black hole space-time. In the latter case, there is a free
angle parameter that makes fine-tuning unnecessary. As a result, small $N$
is compatible with arbitrary direction of the velocity. It is worth
reminding that in the BSW effect, fine-tuning for one particle was mandatory 
\cite{ban}, \cite{prd}. Now, the situation in a sense is opposite: it is
seen from Table 2, that, rather, special conditions are required to avoid
high-energy collisions!

It was already pointed out in Sec. IX of \cite{di} that collisions with high 
$E_{c.m.}$ near two black holes can hint that a similar phenomenon should
occur near more realistic spinning binary black holes. However, as without
special symmetry the analysis of particle motion is too difficult, this
remained as some hope. The results obtained in the present paper can be
considered as partial confirmation of these hopes since we do not use
details of equations of motion and rely on general kinematic reasonings.
Therefore, the general approach under consideration seems to apply to the
more realistic case of rotating black holes as well, although detailed
separate analysis is desirable here.

\section{Acknowledgment}

This work was funded by the subsidy allocated to Kazan Federal University
for the state assignment in the sphere of scientific activities.

\end{document}